\documentclass[epjST]{svjour}

\usepackage{graphics}

\usepackage{amssymb,amsmath,bbm}
\usepackage{hyperref,graphicx}
\newcommand{\bs}[1]{{\boldsymbol{#1}}}
\newcommand{\bk}{\bs{k}}
\newcommand{\br}{\bs{r}}
\newcommand{\bp}{\bs{p}}
\newcommand{\rmd}{\mathrm{d}}

\newcommand{\bra}[1]{\left\langle #1 \right |}
\newcommand{\ket}[1]{\left| #1 \right\rangle}

\newcommand{\avg}[1]{\overline{#1}}
\newcommand{\mv}[1]{\langle #1\rangle}

\newcommand{\nk}{n_\bk} 
\newcommand{\np}{n_\bp} 
\newcommand{\nkpr}{n_{\bk'}} 
\newcommand{\nkzero}{n_{\bk_0}} 
\newcommand{\taus}{\tau_\text{s}}
\newcommand{\tautr}{\tau_\text{tr}}

\begin{document}

\title{Momentum isotropisation in random
  potentials}

\author{T. Plisson\inst{1} \and T. Bourdel\inst{1} \and 
C. A. M\"{u}ller\inst{2}\fnmsep\thanks{\email{cord.mueller@nus.edu.sg}}}

\institute{%
Laboratoire Charles Fabry UMR 8501, Institut d'Optique, CNRS, Univ Paris Sud 11, 2 Avenue Augustin Fresnel, 91127 Palaiseau cedex, France
\and 
Centre for Quantum Technologies, National University of Singapore, 117543 Singapore
}

\abstract{%
When particles are multiply scattered by a random potential,
their momentum distribution becomes isotropic on average. We study 
this quantum dynamics numerically and with a master equation. We show how to measure the
elastic scattering time as well as characteristic isotropisation 
times, which permit to reconstruct the scattering phase function, even in
rather strong disorder.
 } 

\date{\today} 

\maketitle

\section{Introduction} 

These days, ultra-cold atoms permit to study disorder physics in a
quantitative manner. In particular, quantum transport of
non-interacting particles can be studied in 1D \cite{loc1D}, 
2D \cite{diffusion,Labeyrie2012,Jendrzejewski2012b} and 3D \cite{DeMarco,Jendrzejewski2012}. 
One of the latest development, suggested in \cite{Cherroret2012},  consists in 
launching a quasi-monochromatic wave packet 
inside the bulk of a random potential and monitoring the 
single-particle momentum distribution 
\cite{Labeyrie2012,Jendrzejewski2012b}. The long-time evolution
leads to coherent backscattering  (CBS) and
coherent forward scattering (CFS) 
signals, which are linked to weak and strong localization, respectively  
\cite{CFS}.  In this paper, we analytically and numerically analyze the short-time dynamics, which contains valuable
information about disorder strength and scattering
characteristics. In particular, we show how one can 
 not only measure the scattering and transport times, but also reconstruct, by an angular Fourier analysis of
the momentum distribution, the
complete scattering phase function. 

\section{Momentum-distribution dynamics}
\label{me.sec} 

Starting from the momentum distribution $\nk(0)$, we seek to
compute $\nk(t)=\avg{|\psi_\bk(t)|^2}$ at later times $t$ 
under the evolution with $H=H_\text{kin} +V(\br)$. The over-bar $\avg{\{.\}}$ denotes an  
ensemble average over realizations of the random potential $V(\br)$.  

\begin{figure}
\includegraphics[width=\textwidth]{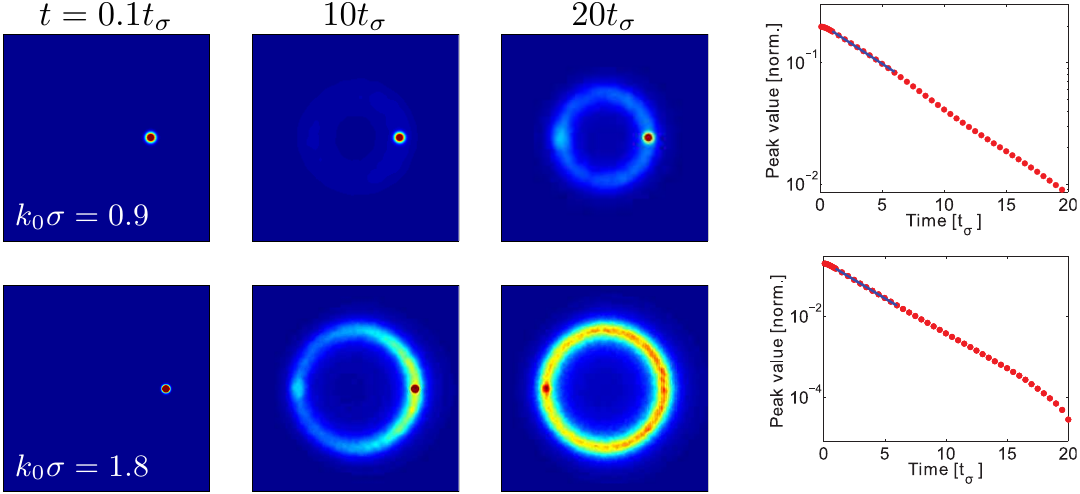}
\caption{Density plots of momentum distributions from the numerical
  simulation. Upper row: Experimental regime of
  \cite{Jendrzejewski2012b}, where $k_0=0.9/\sigma$, $V=0.35E_{\sigma}$, and
  $\Delta k=0.07k_0$. Bottom row: $k_0=1.8/\sigma$, $V=0.7E_{\sigma}$
  and  $\Delta k=0.035k_0$. Parameters are chosen such that the
  weak-disorder ratio $\Delta =
2V^2/(k_0\sigma)^2 \approx 0.3$ of perturbation theory is the same \cite{Kuhn2007}. 
Right panels: the peak value $n_{\bk_0}$ of the initial momentum component 
decreases exponentially, Eq.~\eqref{nkzero}. 
The indicated fits yield the elastic scattering times 
$\gamma_0^{-1}=\taus=6.7t_{\sigma}$ and $2.4t_{\sigma}$, respectively.}
\label{fig1}
\end{figure}

We simulate this evolution numerically by solving the two dimensional (2D) Schr\"{o}\-dinger equation
for matter waves with $H_\text{kin}= \bp^2/2m$ 
using  a finite-difference method. The initial wave function has a
small width $\hbar \Delta k$ around a finite average momentum
$\hbar \bk_0$. 
The random potential $V(\br)$ is an isotropic speckle with
a Gaussian spatial covariance length $\sigma$ and amplitude $V$. The
natural energy and time scales of the system are 
$E_{\sigma}=\hbar^2/m\sigma^2$ and $t_{\sigma}=m\sigma^2/\hbar$.
Figure \ref{fig1} shows numerically calculated  
momentum distributions, averaged over 1000 disorder realizations, for two different 
regimes:   rather isotropic scattering with $k_0\sigma=0.9$ (upper row)
and rather anisotropic scattering $k_0\sigma=1.8$ (bottom
row). 

As the initial peak at $+\bk_0$ decays, due to elastic scattering by
the random potential, other momenta are populated on
the (disorder-broadened) Rayleigh ring of radius $k_0$. 
Eventually, a CBS peak appears at
$-\bk_0$. In the remainder of this
paper, we rather focus on the dynamics that produces the isotropic background. 

From an analytical point of view, since unitary evolution
conserves energy, it is appropriate to study the joint  distribution 
$\nk(E,t)$ of energy and momentum, with marginal $\nk(t) = \int\frac{\rmd E}{2\pi}
\nk(E,t)$.  
The density $\psi\psi^*$ propagates 
according to
\begin{equation}\label{eom}
 \nk(E,t) =  \int\frac{\rmd \omega}{2\pi}
e^{-i\omega t} \sum_{\bk'}\Phi_{\bk\bk'}(E,\omega)
\nkpr(0),   
\end{equation} 
where the intensity propagator $\Phi_{\bk\bk'}(E,\omega)$ 
obeys the integral equation of motion \cite{Kuhn2007,AkkerMon} 
\begin{equation} \label{BS}
[-i\omega + \gamma_{\bk}(E)]\Phi_{\bk\bk'}(E,\omega) = A_{\bk} (E)\left[
 \delta_{\bk\bk'} +
 \sum_{\bp}U_{\bk\bp}(E,\omega)\Phi_{\bp\bk'}(E,\omega)\right].
\end{equation} 
This form is valid for small frequencies $\omega\ll E$ and can thus
describe the dynamics at long enough times $t\gg \hbar/E$.  The spectral function 
$A_{\bk}(E) = 2\pi\bra{\bk}\avg{\delta(E-H)}\ket{\bk}$
is the probability density of a plane-wave state $\bk$ to have energy $E$. 
The vertex $U_{\bk\bp}(E,\omega)$ describes all scattering events 
that couple $\psi$ and $\psi^*$. It is 
related to the elastic scattering rate $\gamma_{\bk}(E)$ 
by the so-called Ward identity \cite{Kuhn2007,Vollhardt1980} 
\begin{equation}\label{gammakEWard}
\gamma_\bk(E) = \sum_{\bp} A_{\bp}(E) U_{\bp\bk}(E,0). 
\end{equation}  
This relation is a variant of the optical theorem, 
and expresses particle-number
conservation. Using $(-i\omega)\Phi_{\bk\bk'}(E,\omega)$ from \eqref{BS} in
\eqref{eom} then yields the equation of motion   
\begin{align} \label{rhokin}
 \partial_t\nk(E,t) = &  A_{\bk}(E)\nk(0)  \delta(t) -
\gamma_{\bk}(E) \nk (E,t) \nonumber \\ & + A_{\bk}(E) \int\frac{\rmd \omega}{2\pi}
e^{-i\omega t}  \sum_{\bp,\bk'}  U_{\bk\bp}(E,\omega)\Phi_{\bp\bk'}(E,\omega) \nkpr(0).
\end{align}

Initially, the populations at $\bk'\neq\bk_0$  
outside the initial wave packet are small. The most noteworthy 
evolution is the decay of $\nkzero$ as described by the first line of 
\eqref{rhokin}, whose solution immediately leads to   
$\nkzero(t) =\theta(t) \left[\int\frac{\rmd E}{2\pi}A_{\bk_0}(E)
  e^{-\gamma_{\bk_0}(E)t}\right] \nkzero(0).$   
This is a superposition of simple exponential decays, weighted by
the spectral density. As function of $E$,  $A_\bk(E)$ is rather
sharply peaked around the
effective excitation energy  $E_\bk$, whereas $\gamma_\bk(E)$ is smooth. Therefore, the dominant
behaviour is  
\begin{equation} \label{nkzero}
\nkzero(t)   =e^{-\gamma_0 t} \nkzero (0)
\end{equation} 
 with the elastic scattering rate 
$ \gamma_0 =  \gamma_{\bk_0}(E_{\bk_0})$. 

The corresponding 
numerical data are shown in the right-hand panels of Fig.~\ref{fig1}. 
For the two regimes of rather strong disorder studied here, we find
$\tau_s=6.5t_{\sigma}$ for the first case and $\tau_s=2.4t_{\sigma}$
for the second. Lowest-order perturbation theory 
  \cite{Kuhn2007,Hartung2008} overestimates the scattering strength and predicts 
  $\tau_s=2.5t_{\sigma}$ and $\tau_s=1.4t_{\sigma}$,
  respectively. 
We have, however, verified that if the simulation is run for $V^2/EE_{\sigma} =
0.005 \ll 1$, the measured value of
$\taus$ agrees with 
  \cite{Kuhn2007,Hartung2008}. 

\section{Diffusive isotropisation}
\label{isotrop.sec}

\subsection{Master equation} 

Equation \eqref{rhokin} describes the
equilibration of populations due to
potential scattering.  Whenever the potential-scattering vertex $U_{\bk\bk'}(E,\omega)$ is
regular for $\omega\to 0$, the singular $\omega$-dependence of $\Phi_{\bk\bk'}(E,\omega)$ dominates the
$\omega$-integral.%
\footnote{$U(\omega)\approx U(0)$ can only be factorised from the
$\omega$-integral in \eqref{rhokin} if it has no singular dependence in
$\omega$. This is true at short times, since the single-scattering term does not depend on frequency at
all and the frequency dependence is smooth for the first 
interference corrections. But the CBS contribution 
develops a diffusion pole at
$\omega=0$ and backscattering around the transport time $\tautr =\tau_1$ (see below). Thus, the
momentum distribution retains a memory of the initial condition and
displays the CBS peak \cite{Cherroret2012}. Even more singular terms appear at
the onset of Anderson localisation and produce the CFS signal
\cite{CFS}. 
These phenomena happen at times $t\gtrsim\tautr$ and much
  longer, beyond the master-equation approach developed here.} 
Using
\eqref{eom} and the Ward identity
\eqref{gammakEWard}, one then arrives at 
a Markovian master equation of the Pauli type: 
\begin{equation} \label{Pauli}
 \partial_t\nk(E,t)  =  \sum_{\bp} U_{\bk\bp}(E) \left [A_\bk(E) 
 \np(E,t) - A_\bp(E)
 \nk(E,t) \right].    
\end{equation} 
Assuming a factorised solution $\nk(E,t) = A_\bk(E) \nk(t)$, 
we then obtain a closed master equation for the momentum distribution,
$
  \partial_t\nk(t)  =  \sum_{\bp} \bar U_{\bk\bp}
  \left[\np(t) - \nk(t)\right]
$, 
with transition rates 
$\bar U_{\bk\bp} =  \int\frac{\rmd
  E}{2\pi}A_{\bk}(E)A_{\bp}(E)U_{\bk\bp}(E)$.   
The spectral functions 
fix the energies
at $E_\bp\approx E_\bk$. Consequently, we can focus on the angular probability density
$n(\theta)$  on the circle
$\bk=k_0(\cos\theta,\sin\theta)$ of elastic
scattering. There, 
scattering from $\theta$ to
$\theta'$ is described by the on-shell
value $U(\theta-\theta')$. 
It is helpful to re-introduce the 
elastic scattering rate $\gamma_0 = \nu_0 \int \rmd\theta U(\theta)$, 
in accordance with \eqref{gammakEWard}; 
 $\nu_0  = \nu(E_{\bk_0})$ denotes the density of states.   
Finally, 
the elastic isotropisation of momentum is described by [see also Eq.~(6.83) of
\cite{spindiff2009}] 
\begin{equation} \label{dotrhotheta}
 \partial_t n(\theta,t)  = \gamma_0 \int_0^{2\pi}\rmd\theta' u(\theta-\theta')
 \left [n (\theta',t) - n(\theta,t) \right],   
\end{equation} 
where  
$u(\beta) = U(\beta) [\int\rmd\theta U(\theta)]^{-1}$ is the
normalised differential cross section of elastic scattering, the 
so-called phase
function. 

 \subsection{Characteristic times} 

The convolution \eqref{dotrhotheta} on the unit circle can
be readily solved by Fourier analysis: the coefficients 
$c_m = \int\rmd\theta e^{im\theta}n(\theta)$
obey the decoupled equations of motion
\begin{equation} \label{dotrhom}
\partial_t c_m(t) = -\gamma_0 (1-u_m)c_m(t)=-\tau_m^{-1}  c_m(t)
 \end{equation}
in terms of the phase-function Fourier components $u_m = \int\rmd\beta
e^{im\beta}u(\beta)$. By virtue of $u_0=1$, the zeroth mode $c_0$
(i.e., the number of particles) is conserved.  
The first harmonic $c_1$ decays exponentially on the transport-time scale   
$\tau_1 = \taus/[1-\mv{\cos\beta}]$, 
with the usual notation $\mv{\cos\beta}= u_1 = \int\rmd\beta
\cos{\beta}u(\beta)$ \cite{Kuhn2007} . 
Higher harmonics decay with their own
times $\tau_m = \taus/ [1-\mv{\cos m\beta}]$. 

Only for a completely
isotropic phase function with $\mv{\cos m\beta}=0$ all times $\tau_m$ are strictly identical to the elastic scattering time $\taus$. 
Things are more interesting for spatially correlated
potentials. Indeed, at the single-scattering approximation, the
Gaussian correlator $U_{\bk\bp}^{(1)}\propto
\exp\{-\sigma^2(\bk-\bp)^2/2\}$ has the phase function  
$u^{(1)}(\beta) = \exp\{k^2\sigma^2\cos\beta\}/[2\pi I_0(k^2\sigma^2)]$
where $I_0(\cdot)$ is the modified Bessel function. 
The corresponding Fourier coefficients 
\begin{equation}\label{umgauss} 
u_m ^{(1)} = I_m(k^2\sigma^2)/I_0(k^2\sigma^2)
\end{equation}  
decrease as a function of the
order $m$, such that the decay rates $\tau_m^{-1}$ increase.  
The complete isotropisation dynamics can only be described by the entire
  set of times $\tau_m$. In general, higher-order modes decay more
  quickly, giving way to the fundamental, diffusion mode with the longest
  lifetime  $\tautr=\tau_1$.

\begin{figure}
\centering
\includegraphics[width=0.31\textwidth]{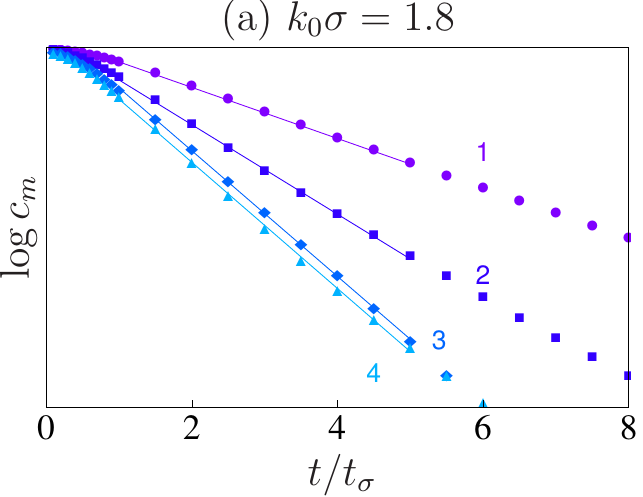}
\includegraphics[width=0.33\textwidth]{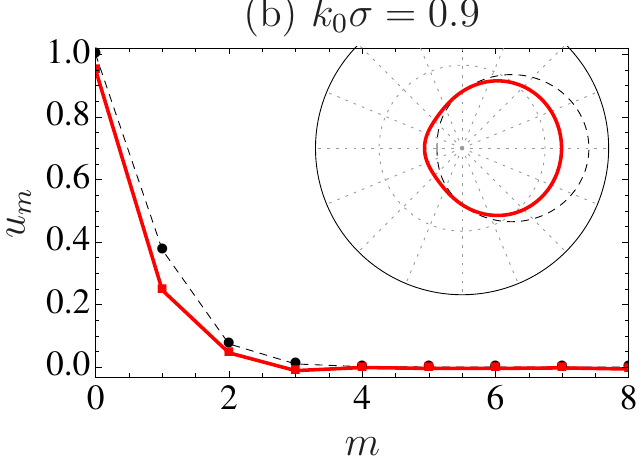}  
\includegraphics[width=0.33\textwidth]{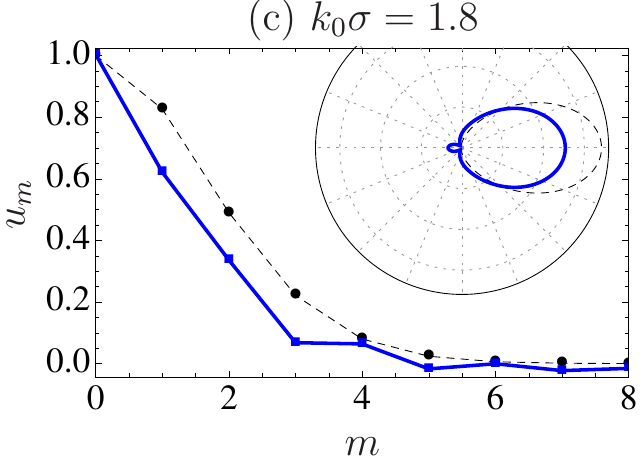}
\caption{(a) Fourier coefficients of the momentum distribution on the elastic scattering circle, up to
  $m=4$, as function of time, for the case $k_0\sigma=1.8$. 
   Straight lines
  are fits to the exponential decay, Eq.~\eqref{dotrhom}, for $t_{\sigma}<t<5 t_{\sigma}$. 
  (b) and (c) Fourier coefficients $u_m$ of the scattering phase function, 
  as measured
from the numerical data (full red/blue), and at
the single-scattering approximation, Eq.~\eqref{umgauss} (dashed black).  
The insets show the reconstructed phase functions, using moments up to $m=4$.}  
\label{fig2}
\end{figure}

In Fig.~\ref{fig2}(a), we show the time evolution of the first four
Fourier coefficients $c_m$ of the angular distribution in the case
$k_0\sigma=1.8$, as obtained from the numerical simulation
data shown in Fig.~\ref{fig1} by a radial integration over a small
interval around $|\bk|=k_0$.

\subsection{Reconstruction of the phase function} 

From the measured decay of $c_m(t)$, one can compute the
Fourier coefficients $u_m$
via Eq.~\eqref{dotrhom}, and thus reconstruct the phase function $u(\beta)$. 
Fig.~\ref{fig2}(b) and (c) show the $u_m$
as measured from the numerical data, together with the single-scattering
approximation \eqref{umgauss}. In the first case $k_0\sigma=0.9$, the
higher-order Fourier coefficients of the momentum distribution are
very small. Consequently, the scattering phase function, shown in the
inset of Fig.~\ref{fig2}(b), is rather isotropic, as expected for  
$k_0\sigma<1$.  In the second case $k_0\sigma=1.8$, the scattering is
much more pronounced in the forward direction, as expected for 
$k_0\sigma\gg1$. 

Compared to the bare single-scattering phase functions (in dashed),
the full reconstructed phase functions show a tendency towards
enhanced backscattering. We may attribute this to weak
localization corrections associated with short scattering paths, 
which are only known for isotropic scatterers \cite{Eckert2012}.

\section{Conclusion}
\label{conclusion.sec} 

We have analyzed the momentum isotropisation of a quasi-monochromatic
wave packet inside a spatially correlated random potential. 
Results from the numerical calculation for an optical speckle
potential are well reproduced by a master-equation
approach, even in regimes of strong disorder. We show how to
reconstruct the phase function of elastic scattering from the measured
exponential decay times of the angular components 
of the momentum distribution. This method is directly applicable in
present-day experiments \cite{Labeyrie2012,Jendrzejewski2012b}, and
can serve as an  \textit{in situ} calibration of random potentials,
for instance when investigating the impact of disorder in the presence
of interactions \cite{Allard2012,Beeler2012,Krinner2012}.  

\paragraph{Acknowledgements: } 

T.P. is supported by IXBLUE. C.M.\ acknowledges financial support from Fondation des Sciences
Math\'ematiques de Paris (FSMP) within the programme ``Disordered
Quantum Systems'' at Institut Henri Poincar\'e. LCFIO is member of
IFRAF and supported by CNRS, RTRA triangle de la physique,
ANR-08-blan-0016-01, Isense. 


\end{document}